\documentclass[pdflatex,sn-nature]{sn-jnl}

\usepackage{graphicx}%
\usepackage{multirow}%
\usepackage{amsmath,amssymb,amsfonts,bm}%
\usepackage{amsthm}%
\usepackage{mathrsfs}%
\usepackage[title]{appendix}%
\usepackage{xcolor}%
\usepackage{textcomp}%
\usepackage{manyfoot}%
\usepackage{booktabs}%
\usepackage{algorithm}%
\usepackage{algorithmicx}%
\usepackage{algpseudocode}%
\usepackage{listings}%
\usepackage{upgreek}
\usepackage{xr}

\usepackage[normalem]{ulem}

\makeatletter

\newcommand*{\addFileDependency}[1]{
\typeout{(#1)}%
\@addtofilelist{#1}
\IfFileExists{#1}{}{\typeout{No file #1.}}
}\makeatother

\newcommand*{\myexternaldocument}[1]{%
\externaldocument{#1}%
\addFileDependency{#1.tex}%
\addFileDependency{#1.aux}%
}
\myexternaldocument{SI}
\myexternaldocument{ExtendedData}

\begin{document}

\title[Article Title]{Deep Strong light--matter Coupling in 3D Kane Fermions}

\author*[1,2]{\fnm{Dmitriy} \sur{Yavorskiy}}\email{dmitriy.yavorskiy@fuw.edu.pl}
\author[3]{\fnm{David} \sur{Hagenm\"uller}}
\author[1]{\fnm{Noureddine} \sur{Charrouj}}
\author[1,2]{\fnm{Yurii} \sur{Ivonyak}}
\author[4]{\fnm{Alexander} \sur{Kazakov}}
\author[5]{\fnm{Yanko} \sur{Todorov}}
\author[1,2]{\fnm{Wojciech} \sur{Knap}}
\author*[1,2]{\fnm{Marcin} \sur{Białek}}\email{marcin.bialek@unipress.waw.pl}

\affil[1]{\orgdiv{CENTERA Labs, Institute of High Pressure Physics}, \orgname{Polish Academy of Sciences}, \orgaddress{\city{Warsaw}, \postcode{01-142}, \country{Poland}}}
\affil[2]{\orgdiv{CENTERA, CEZAMAT}, \orgname{Warsaw University of Technology}, \orgaddress{\city{Warsaw}, \postcode{02-822}, \country{Poland}}}
\affil[3]{\orgdiv{Institut de Physique et Chimie des Matériaux de Strasbourg, UMR 7504}, \orgname{Universit\'{e} de Strasbourg and CNRS}, \orgaddress{\city{Strasbourg}, \postcode{67000}, \country{France}}}
\affil[4]{\orgdiv{International Research Centre MagTop, Institute of Physics}, \orgname{Polish Academy of Sciences}, \orgaddress{\city{Warsaw}, \postcode{02-668}, \country{Poland}}}
\affil[5]{\orgdiv{Laboratoire de Physique et d'\'{E}tude des Mat\'{e}riaux, LPEM, UMR 8213, ESPCI Paris}, \orgname{Universit\'{e} PSL}, \orgaddress{\city{Paris}, \postcode{F-75005}, \country{France}}}

\newpage

\abstract{Deep strong light-matter coupling represents an extreme non-perturbative regime of quantum electrodynamics, in which the interaction strength exceeds the bare frequencies of the uncoupled systems. The ground state features strong quantum correlations between photons and matter excitations, and new cavity-driven phase transitions are expected to occur. 
Whether a superradiant quantum phase transition, marked by spontaneous dipole ordering and photon condensation, is possible has remained a long-standing and controversial question. Such phenomena have been proposed to arise in exotic electronic systems hosting Dirac and Kane fermions, owing to the formal absence of an $A^2$ term in their low-energy Hamiltonian. Here we exploit the ultralow effective mass of Kane fermions to realise Landau polaritons in a bulk mercury cadmium telluride layer coupled to a Fabry–Perot resonator. Using thermally tunable carrier density, we continuously tune the coupling from the weak to the deep-strong regime, achieving a record normalised coupling ratio exceeding 1.6 above room temperature. The measured polariton spectra are in excellent agreement with a rigorous, gauge-invariant microscopic theory. Despite the nonlinear Landau level structure of relativistic Kane fermions, we show that a diamagnetic $A^2$ term naturally emerges and precludes a superradiant phase transition. These results resolve the long-standing controversy surrounding cavity quantum electrodynamics of relativistic-like matter systems, extend deep-strong-coupling physics to Kane fermions, and open new opportunities for polaritonic semiconductor devices operating in extreme light-matter coupling regimes.}


\maketitle

\section{Introduction}\label{sec:intro}

Strong light--matter coupling in cavity quantum electrodynamics (QED) arises when coherent energy exchange between matter and photonic modes exceeds dissipation rates, leading to hybrid light--matter excitations known as polaritons~\cite{HJKimble:PhysScr1998,JJHopfield:PR1958}. In the ultrastrong coupling regime, where the vacuum Rabi frequency $\Omega$ becomes a significant fraction of the bare transition frequency $\omega$, the rotating-wave approximation breaks down, and the ground state acquires a squeezed-vacuum character~\cite{Ciuti2005,kockum2019ultrastrong,Forn-Diaz2019}. In the deep-strong-coupling regime ($\Omega/\omega > 1$), light--matter interactions become fully non-perturbative and generate large ground-state photon populations~\cite{Ciuti2005,Forn-Diaz2019,kockum2019ultrastrong}. These regimes lie at the heart of polaritonics, a rapidly growing field exploring how cavity QED can reshape material properties~\cite{Garcia-Vidal2021,Bloch2022StronglyCorrelated,Bretscher2026FluctuationEngineering}, including phase transitions~\cite{Jarc2023CavityMediatedMIT,Xu2026VacuumDressed,Keren2026CavityAlteredSuperconductivity}, chemical reactivity~\cite{Hutchison2012,ThomasGS2019,Garcia-Vidal2021}, and charge~\cite{orgiu_conductivity_2015,Appugliese2022Breakdown} and energy transport~\cite{Andrews_2000,Zhong2017}. 

The superradiant quantum phase transition is a paradigmatic illustration of how polaritonics could push the frontier in quantum materials research. Originally predicted within the Dicke model, describing a collection of two-level systems coupled to an electromagnetic cavity mode~\cite{RHDicke:PhysRev1954}, it arises when the light-matter coupling strength exceeds a critical threshold~\cite{HeppLieb1973}. In the superradiant phase, which typically lies in the deep-strong-coupling regime, the ground state exhibits both photon condensation and macroscopic matter polarization~\cite{EmaryPRL2003}. In realistic systems, however, the diamagnetic $\mathbf{A}^2$ term imposes stringent constrains on such a transition~\cite{Andolina2020,Guerci2020}, which translates in no-go theorems~\cite{Rzazewski1975,NatafCiuti2010,Andolina2019}. About a decade ago, it was theoretically proposed that the absence of an $\mathbf{A}^2$ term in the minimally coupled low-energy Dirac model of graphene, owing to the linear dependence of the Dirac Hamiltonian on momentum, could circumvent these no-go theorems and enable the superradiant phase transition~\cite{Hagenmuller2012}. This prediction was subsequently challenged because an effective $\mathbf{A}^2$ term should be dynamically generated to ensure gauge invariance~\cite{Chirolli:PRL2012,Andolina2019}. Kane fermions, extending the concept of Dirac fermions, provide an ideal platform for investigating this phenomenon. Here, we achieve the deep-strong-coupling regime between the photonic modes of a Fabry-Perot resonator and the cyclotron-resonance (CR) transitions between Landau Levels (LLs) of 3D Kane fermions.  

Previous experimental realisations, such as Landau~\cite{GScalari:Science2012} or magnetoplasmon~\cite{SRajabali:Natphot2021,SRajabali:NatCom2022,LHalearxiv2025} polaritons, were reported mainly with 2D electron gas from gallium arsenide (GaAs) quantum wells with parabolic band dispersion, where the effect of the $\mathbf{A}^2$ terms is well-known \cite{DHagenmuller:PRB2010}. This platform has enabled engineering of polaritonic nonlinearities~\cite{JMornhinweg:PRL2021}, modifications of magnetotransport~\cite{NBartolo:PRB2018,GParaviciniBagliani:Natphys2019}, quantum Hall response~\cite{CCiuti:PRB2021,FAppugliese:Science2022}, and excitation spectra of quantum materials~\cite{JEnknerNat2025,Bacciconi2025GravitonPolaritonsPRX}. Deeply subwavelength metasurface resonators have enabled record coupling ratios up to $\Omega/\omega=1.4$~\cite{ABayer:Nanoletters2017,Baydin2025DeepStrongPbTe}, whereas Fabry-Perot~\cite{EMavrona:ACSPhot2021} and photonic crystal cavities~\cite{QZhang:Natphys2016,FTay:Natcommun2025} offer weaker confinement but substantially higher quality factors, opening regimes where high cooperativity coexists with ultrastrong coupling~\cite{QZhang:Natphys2016,LXinwei:NatPhot2018,FTay:Natcommun2025}. 

\begin{figure}[ht!]
    \includegraphics[width=1.\textwidth]
    {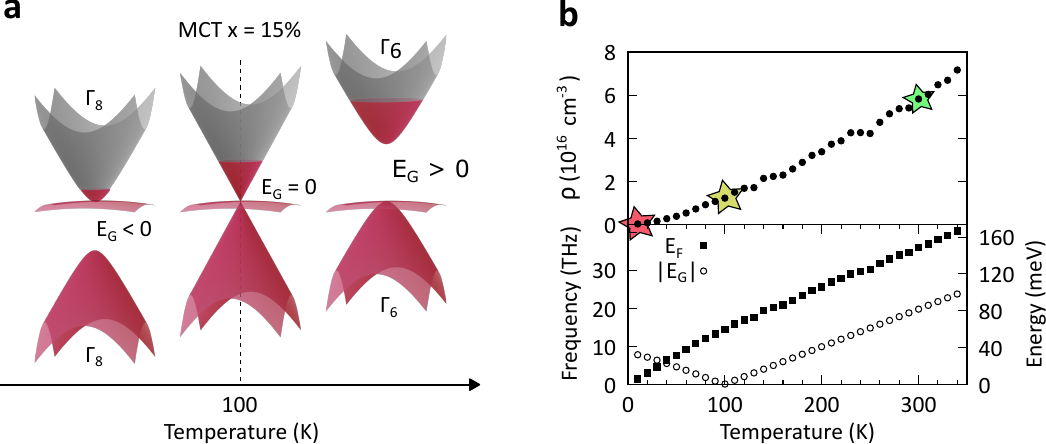}
        \caption{\textbf{Kane fermions in MCT.}
(a) Temperature evolution of the electronic band structure from an inverted semimetallic phase ($E_{\mathrm{G}}<0$) at low temperatures, to a normal semiconducting phase ($E_{\mathrm{G}}>0$) at high temperatures. At the critical temperature $T_c\approx100$~K (for $x=0.15$), the band structure becomes gapless. At this point, the bands exhibit linear momentum dispersion and host massless Kane fermions.
With increasing temperature, the Fermi level rises into the conduction band (red).
(b) Top, temperature dependence of the electron density in MCT; coloured stars mark the temperatures corresponding to the reflection spectra shown in Fig.~\ref{fig:fig2}. Bottom, temperature dependence of the Fermi energy ($E_F$) and band gap ($|E_G|$) (both shown in meV and converted to THz).}
\label{fig:fig1}
\end{figure}

Experimental realisations of extreme coupling in bulk three-dimensional electronic systems remain scarce, whereas narrow-gap semiconductors with relativistic-like dispersion~\cite{WZawadzkiAdvPhys1974} offer a promising route. Mercury cadmium telluride (Hg$_{1-x}$Cd$_x$Te, MCT) is a paradigmatic example, whose band structure can be tuned by composition, temperature, or pressure through a topological phase transition via a gapless critical point~\cite{MOrlita:NatPhys2014,FTeppe:NatCom2016}. Near this point, the conical dispersion at the $\Gamma$ point intersects a nearly flat heavy-hole band (Fig.~\ref{fig:fig1}a), yielding a bulk analogue of relativistic quasiparticles. Those MCT 3D Kane fermions~\cite{EvanOKane:1957} with linear dispersion exhibit ultralight effective mass and unconventional collective charge dynamics~\cite{ACharnukha:ScienceAdvances2019}, enabling efficient coupling with terahertz (THz) fields. In magnetic fields, the strong magneto-optical response and large effective $g$-factor support electric-dipole-allowed spin-split inter- and intra-LL transitions naturally in the THz range~\cite{MOrlita:NatPhys2014,FTeppe:NatCom2016}. 

The experimental demonstration of the highly non-perturbative light-matter interaction reported here establishes MCT as a tunable material for THz polaritonics. We use MCT with $x=0.15$, which hosts Kane fermions near the critical temperature $T_c\approx100$~K. As temperature increases, the band gap evolves (Fig.\ \ref{fig:fig1}b), and carriers are thermally excited into the conduction band, raising the Fermi level~\cite{YNemirovsky:JAP1979,GLHansen:JAP1983,Seiler1990,Moldavskaya24,Yahniu25}. Increasing electron concentration provides a route to tune the collective light--matter coupling with temperature. Combined with the ultralight effective mass and intrinsic Fabry-Perot cavity modes in our samples, this enables Landau polaritons to be continuously tunable from the weak- to the deep-strong-coupling regime. We reach a record coupling ratio $\eta>1.6$ beyond room temperature. Combined THz magneto-spectroscopy and microscopic modelling reveal exceptionally large collective coupling without subwavelength metasurfaces. Besides providing a scalable route to extreme regimes of light--matter interactions, this platform offers an experimental test of the presence of an effective $\mathbf{A}^2$ term in the low-energy model, and of whether it may allow a superradiant phase transition. We find that it does not. Our gauge-invariant microscopic quantum model with proper low-energy truncation is in excellent agreement with the experimental results, highlighting the importance of consistent truncation schemes in cavity QED.

\section{Samples and experiment}

We investigate MCT/CdTe heterostructures with a $h=4~\mu$m-thick molecular beam epitaxial layer of MCT with $x = 0.15$. We fabricate the photonic cavity by bonding the MCT side to Cu/Au-coated laminate using a dielectric layer (Fig.~\ref{fig:fig2}a). By chemical etching, we remove the GaAs substrate, and the resulting structure creates a Fabry-Perot cavity. Details of the sample design and fabrication are provided in the Methods sec.\ Sample Preparation.

We use the transfer-matrix method to calculate the spatial distributions of the electric field in the sample (Supplementary Information Sec.\ S2). The model incorporates temperature-dependent 3D conductivity in MCT, and a temperature-dependent harmonic oscillators introducing the phonon modes.
These calculations indicate that, at zero magnetic field and low temperatures (low carrier concentration in the MCT layer), the structure supports two resonances in our spectral range, below 3.5~THz (Fig.~\ref{fig:fig2}b). The lowest mode $M_1^0$, near 1~THz, is a photonic $22$~$\mu m =L=\lambda_1/4$ mode. Such quarter-wave modes form due to asymmetric boundary conditions, with a node at the metal-dielectric interface and an anti-node at the vacuum-CdTe interface. The second mode $M_2^0$, at about 3~THz, arises from the resonant strong coupling between the $L=3\lambda_2/4$ FP mode and the MCT phonon mode (Supporting Information Sec.\ S2, Fig.\ S4).

The reflectivity of the system as a function of magnetic field $B$ and temperature $T$ is probed experimentally by a commercial THz time-domain spectrometer. Linearly-polarised THz pulses are incident at $8^\circ$ in the Faraday geometry. The presented frequency-domain spectra show the amplitude of the Fourier transform of time-domain reflection traces, i.e.\ these are electric field spectra. They are normalised to the reflection spectrum of the metallic mirror holder. Further experimental details are provided in the Methods, Sec.~Experimental setup.

\section{Zero magnetic field spectra}

\begin{figure}[ht!]
    \includegraphics[width=1.\textwidth]{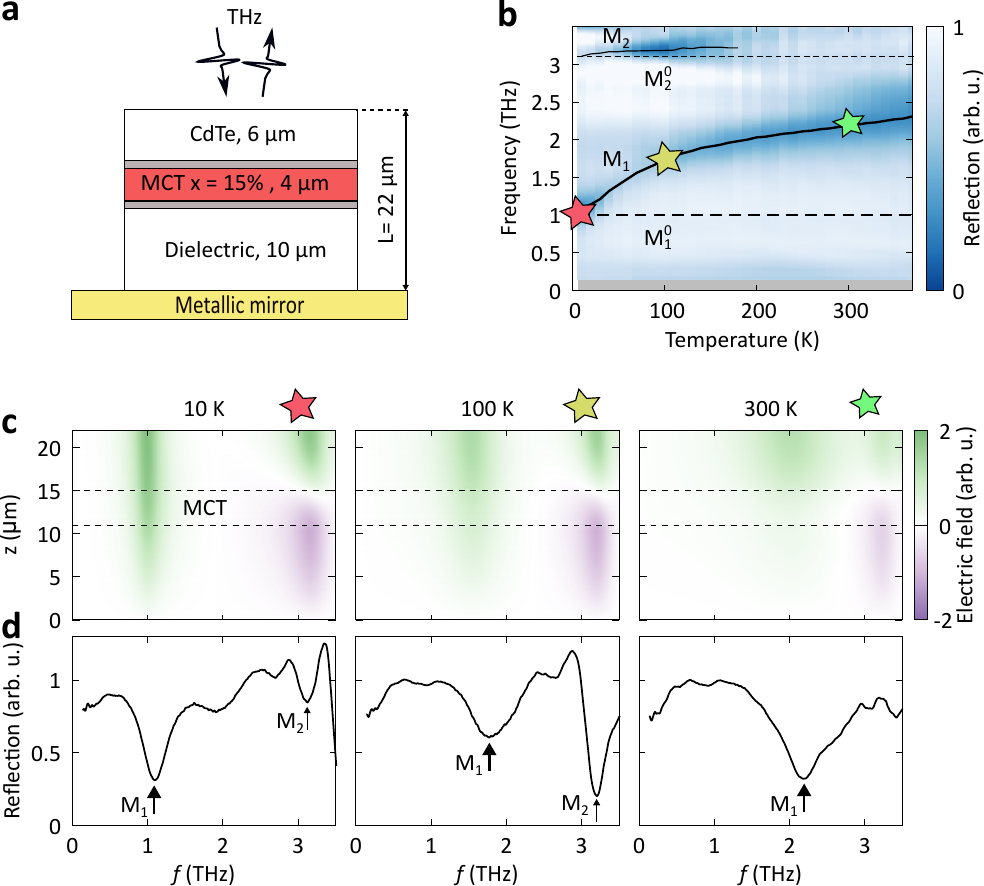}
        \caption{\textbf{Sample and cavity modes.} 
(a) Heterostructure schematic indicates the active MCT layer as the red block, which is sandwiched by $\sim1~\mu$m-thick MCT layers with Cd composition continuously graded from 45\% to 15\% (grey blocks). We probe the sample using THz-TDS in reflection mode. 
(b) Temperature evolution of M$_1$ and M$_2$ modes in reflection spectra (solid lines). M$_1^0$ and M$_2^0$ denote the modes in the absence of carriers in the MCT layer (dashed lines).   
(c) Calculated real part of the electric field profile in the cavity at zero magnetic field and selected temperatures 
(d) Reflection spectra at the same selected temperatures. Arrows mark M$_1$ and M$_2$ modes.}
	\label{fig:fig2}
\end{figure}

In our samples, at low temperatures, MCT has an electron concentration of approximately $2\times10^{14}\,\mathrm{cm^{-3}}$ and a mobility $\sim10^{6}\, \mathrm{cm^2/Vs}$ (Supplementary Information, Sec.\ S6). With rising temperature, electron density increases by two orders of magnitude above room temperature (Fig.\ \ref{fig:fig1}b). This evolution of the carrier density leads to the deep strong coupling of Landau polaritons, reported further as a function of the magnetic field. However, the evolution of electron concentration with temperature also strongly affects the Fabry-Perot resonances at zero magnetic field. 


Figure~\ref{fig:fig2}b presents the temperature evolution of reflection spectra at zero magnetic field (Fig.\ \ref{fig:fig2}d shows spectra for three selected temperatures).
We observe a pronounced resonance, referred later to as M$_1$, across the entire temperature range, rising to higher frequencies with temperature, from 1~THz at 10~K to 2.3~THz at 370~K. The frequency of M$_1$ increases due to increasing carrier density in the MCT layer.
A similar effect was reported in THz plasmonic cavities based on the narrow-gap semiconductor InSb~\cite{Aupiais2023Ultrasmall}. As the carrier concentration increases, the MCT layer becomes an increasingly effective screen for electromagnetic waves near 1.0 THz. As a result, at high temperatures, M$_1$ is pushed out of the MCT layer, reducing the effective mode length. Above about 100~K, the frequency of M$_1$ rises more slowly, because further changes in carrier density no longer lead to significant modifications of the effective mode length. This experimental behaviour is fully consistent with transfer-matrix calculations, which show that at low carrier concentrations, MCT is nearly transparent at M$_1$ frequency (Fig.~\ref{fig:fig2}c, 10~K). The progressive increase of the screening effect can be observed in the field maps for 100~K and 300~K, whereas at 300~K the field is very well localised in the CdTe layer (Fig.~\ref{fig:fig2}c, 300~K). As shown further, this screening effect is directly related to the $\mathbf{A}^2$ term in the light-matter interaction Hamiltonian. 

Above 3 THz, we observe the M$_2$ mode of a nearly temperature-independent frequency (Fig.\ \ref{fig:fig2}b). This is because, even at higher temperatures, the carrier concentration in the MCT layer is still too small to screen electromagnetic fields near 3 THz.
Transfer-matrix calculations show that M$_2$ has a node of the electric field positioned in the MCT layer, which strongly reduce its coupling strength to the CR. 

\section{Landau levels of bulk Kane fermions}

\begin{figure*}[ht!]
     \centering
         \includegraphics[width=0.98\textwidth]{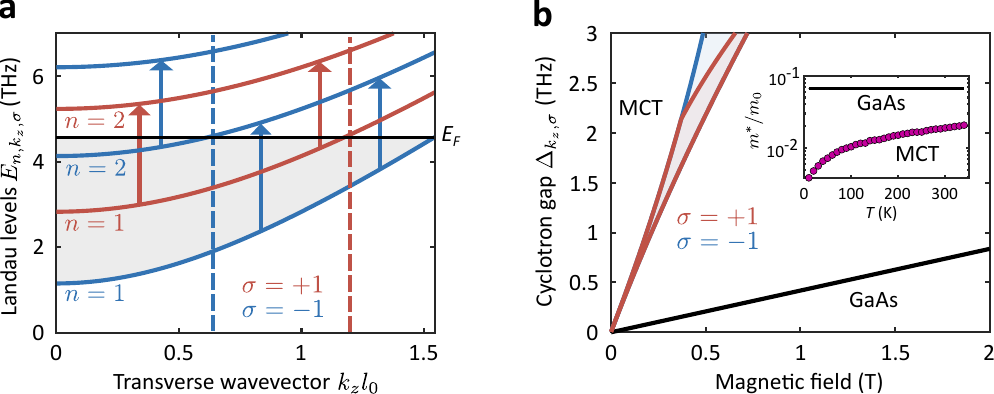}
	\caption{\textbf{LL dispersion and cyclotron transitions in MCT.} 
(a) Energy dispersion of the conduction-band LL $E_{n,k_{z},\sigma}$ at $B=0.45\,$T and $T=30\,$K (electron density $\rho=1.48\times10^{15}\,$cm$^{-3}$). The Fermi energy $E_{\mathrm{F}}$ is shown as a horizontal black line. Filled states lie within the shaded area. Blue (red) curves denote the $\sigma=-1$ ($\sigma=+1$) sector. Dipole-allowed, $\sigma$-conserving transitions between the highest occupied and lowest empty LLs are marked by arrows. As $k_{z}$ moves from left to right across the vertical blue (red) dotted line, the filling factors $\nu_{-}$ ($\nu_{+}$) decrease from $2$ ($1$) to $1$ ($0$).
(b) Cyclotron gap $\Delta_{k_{z},\sigma}$ between the last filled and first empty LLs as a function of magnetic field $B$, forming quasi-continuous bands in the $\sigma=-1$ (blue shaded region) and $\sigma=+1$ (red shaded region) sectors at $T=30\,$K. For comparison, the black line indicates the CR frequency $eB/m_{\mathrm{GaAs}}$ of electrons in the conduction band of GaAs, with $m_{\mathrm{GaAs}}=0.067\,m_{0}$. Inset: temperature dependence of the effective mass ratio $m^{*}/m_{0}$ for MCT electrons (magenta dots) as compared to $m_{\mathrm{GaAs}}/m_{0}$ (horizontal line). All energies are converted to THz.}
	\label{fig:fig11}
\end{figure*}

Our description of electrons in the MCT layer is based on the standard Kane model~\cite{EvanOKane:1957}, as detailed in Ref.~\cite{MOrlita:NatPhys2014}. Retaining only the terms linear in the wavevector and neglecting the $\Gamma_{7}$ band, which lies $\sim1$~eV away from the low-energy sector of interest, the model contains three energy bands (Fig.~\ref{fig:fig1}a), each doubly degenerate due to time-reversal symmetry. In the presence of a static magnetic field $B$, electron bands split into non-equidistant LLs dispersing with the wavevector component along the cavity axis (transverse wavevector $k_{z}$). In this work, we focus on the conduction band, as the transitions between LLs within the valence band are Pauli-blocked in the studied temperature range. Moreover, transitions from either the valence band or the flat band to the conduction band occur at energies much higher than the cavity modes and therefore contribute only marginally to the spectra. The LL energies in the conduction band read
\begin{align}
\label{dispersion_LL}
E_{n,k_{z},\sigma}
&= \frac{E_{\textrm{G}}}{2}
+ \sqrt{\frac{E^{2}_{\textrm{G}}}{4}
+ \left(\frac{\hbar v}{l_{0}}\right)^2 \left(2n-1+\frac{\sigma}{2}\right)
+ \left(\hbar v k_{z}\right)^2},
\end{align}
where $n$ is the LL index (a non-negative integer), $l_{0}=\sqrt{\hbar/(e B)}$ is the magnetic length. The quantum number $\sigma=\pm 1$ arises from the breaking of time-reversal symmetry by the applied magnetic field and can be interpreted as the spin projection along the field direction~\cite{MOrlita:NatPhys2014}. The Kane velocity is taken to be $v \approx 10^{6}\,\mathrm{m/s}$, while the band gap energy is $E_{\textrm{G}}$.  The LL dispersion along $k_{z}$ is shown in Fig.~\ref{fig:fig11}a for a magnetic field $B=0.45\,\mathrm{T}$, in the two sectors $\sigma=\pm 1$ for a fixed electron density at $T=30\,$K obtained by fitting the experimental data with a microscopic model at zero magnetic field, as explained below.  
Filled states below the Fermi energy $E_{\textrm{F}}$ lie within the shaded area. 
Owing to the three-dimensional character of the MCT layer and the non-equidistant LL spectrum, dipole-allowed transitions between the highest occupied level $n=\nu_{\sigma} (k_{z})$ and the lowest empty level $n=\nu_{\sigma}(k_{z})+1$, with transition energy $\Delta_{k_{z},\sigma}=E_{\nu_{\sigma}+1,k_{z},\sigma}-E_{\nu_{\sigma},k_{z},\sigma}$, acquire a finite dispersion along $k_{z}$ (Fig.~\ref{fig:fig11}b). These cyclotron transitions between consecutive LLs therefore form a quasi-continuous band when all allowed $k_{z}$ values are considered (shaded regions in Fig.~\ref{fig:fig11}b). This behaviour contrasts with the linear $B$-field dependence of CR frequency $eB/m_{\mathrm{GaAs}}$, observed in GaAs, where equally spaced LLs remain independent of layer thickness owing to the parabolic $k_z$ dispersion. As shown in the next section, however, the collective character of the light-matter coupling in MCT effectively gives rise to a single ``bright mode'' that couples to the cavity mode. In the low magnetic field regime, $l_{0} \gg \hbar/\sqrt{m^{*}E_{\textrm{F}}}$, the $n$-dependence in Eq.~\eqref{dispersion_LL} becomes negligible, yielding a degenerate CR composed of contributions from many LLs. In this regime, the transition frequency increases linearly with magnetic field, i.e., $\Delta_{k_{z},\sigma} \sim e B/m^{*}$, where the effective mass of the MCT electrons reads $m^{*}=\left|E_{\textrm{F}}-(E_{\textrm{G}}/2)\right|/v^{2}$. This value can be compared to the effective mass $m_{\textrm{GaAs}}=0.067\,m_{0}$, where $m_{0}$ is the free-electron mass.
The inset of Fig.~\ref{fig:fig11}b displays the effective mass ratio. Increase of effective mass is caused by the temperature evolution of the band structure and the rise in Fermi energy. 
At low temperature ($T\sim 10\,\mathrm{K}$), the effective mass in MCT is almost two orders of magnitude smaller than in GaAs, a key factor enabling the exceptionally large light-matter coupling ratios discussed below. 

\section{Deep strong coupling of Kane fermions}

\begin{figure*}[h!]
     \centering
         \includegraphics[width=1.0\textwidth]{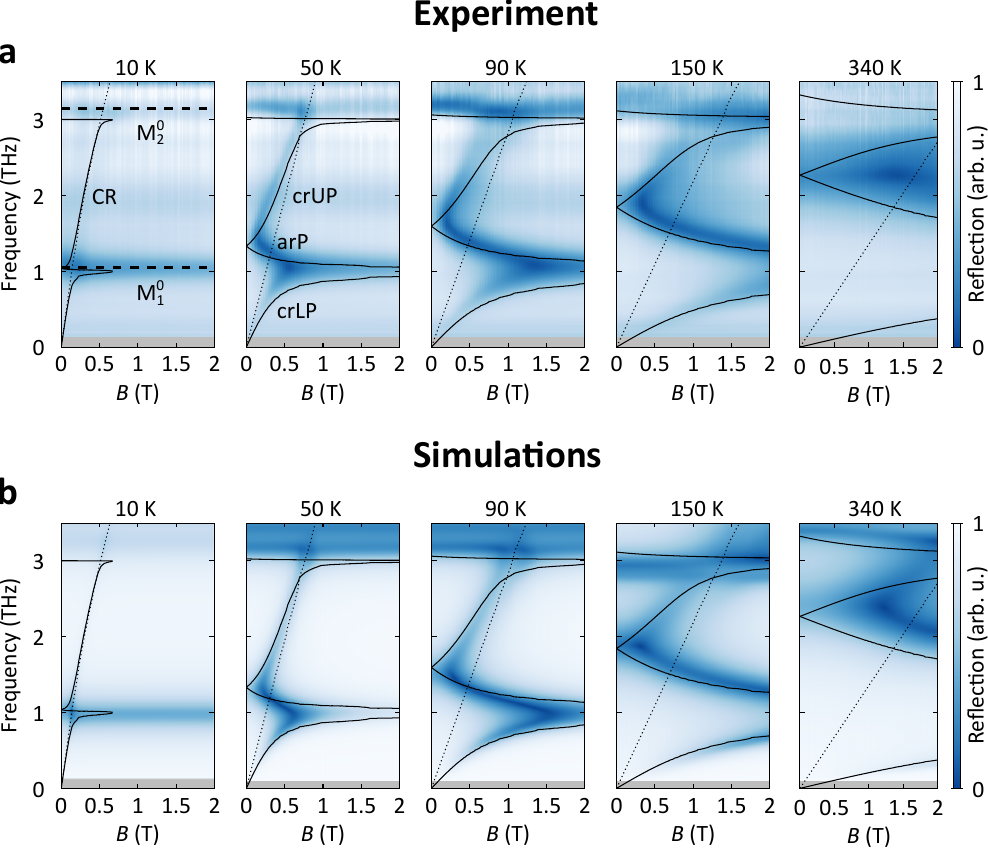}
\caption{\textbf{Reflection spectra.} (a) Reflection 
as a function of magnetic field at selected temperatures. Dotted lines indicate the CR. At 10~K, two horizontal dashed lines mark photonic modes M$_1^0$ and M$_2^0$. Continuous lines are eigenvalues of Eq.\ \eqref{H_total_expanded3}, which are labelled in the 50~K segment---the upper and lower co-rotating polaritons as crUP and crLP, respectively, and the anti-rotating polariton as aRP.
(b) Simulated spectra based on the transfer-matrix model. The lines are identical to those in panel (a).
}
	\label{fig:B5T}
\end{figure*}

Our main results are displayed in Fig.~\ref{fig:B5T}a, which shows the magnetic-field dependence of the reflection spectra measured at different temperatures. At 10~K, we observe the CR with an almost linear dependence on magnetic field. Modes M$_1^0$ and M$_2^0$ are magnetic-field independent and are indicated by black dashed lines. Raw time-domain traces, as well as data for additional temperatures, are available in the Supplementary Information, Sec.\ S3 and S4.

At elevated temperatures, the increased free-electron concentration enhances the coupling strength $\Omega_1$ between the cyclotron resonance (CR) and the M$_1$ cavity mode.
The polarisation degeneracy of the M$_1$ mode in a circular geometry is lifted by its coupling to the CR.
The co-rotating component of the cavity field couples strongly to the CR, giving rise to upper and lower co-rotating polaritons (crUP and crLP).
In contrast, the anti-rotating component couples weakly, leading to the Bloch-Siegert shift~\cite{LXinwei:NatPhot2018}.
Together, these three modes form the characteristic zigzag-shaped Landau-polariton dispersion.

Previous observation with GaAs systems required cavities with narrow linewidths ~\cite{LXinwei:NatPhot2018}. Here, we observe a very strong effect for high electronic density, even though the polariton resonances are much broader. This qualitatively attests to the highly non-perturbative regime of light-matter interaction observed here.
An interesting observation from the data is that, as the magnetic field is increased, the crUP mode asymptotically approaches the unscreened Fabry-Perot mode M$_1^0$ at 1~THz. This is further confirmed by cavity field maps (Supplementary Information, Sec.\ 2 Fig.\ S5).

The CR-M$_2$ interaction $\Omega_2$ also increases, although it remains much weaker than the coupling to M$_1$.
That is caused by a small overlap of M$_2$ electric field with the MCT layer.
Above 200~K, the Landau polariton associated with the M$_2$ mode is pushed outside the studied magnetic field range. 

We simulate the reflection spectra (Fig.\ \ref{fig:B5T}b) using the transfer matrix method, assuming normal incidence on the stack of dielectric layers (Fig.\ \ref{fig:fig2}a). To describe the magneto-optical properties of the MCT layer, we use the gyrotropic dielectric tensor \cite{MIURA2011256}. The transfer matrix method uses the same parameters as the microscopic model (Sec.\ \ref{microscopic_model}), namely the same carrier density (Fig.\ \ref{fig:fig1}b) and the CR frequency (Fig.\ \ref{fig:fig11}b).
It incorporates losses and broadening through the imaginary part of the dielectric functions of layers.
Full details of model parameters are provided in the Supplementary Information, Sec.\ 2.  
Moreover, this method evaluates the electric field distribution in the cavity as a function of the frequency, magnetic field, and temperature.


\vspace{0.5cm}

\section{Microscopic quantum model}
\label{microscopic_model}
To elucidate the coupling of relativistic LLs in MCT to the cavity vacuum field, we developed a microscopic quantum model including the first two open cavity modes of the structure. We model the bare cavity modes as $\lambda_{1}/4$ and $3\lambda_{2}/4$ resonances, M$_1^0$ and M$_2^0$, respectively. The hybrid phonon--photon character of the latter is taken into account through a re-normalised $M_2^0$ frequency and a higher refractive index.
Because the MCT layer thickness $h\simeq4\,\mu$m is much smaller than the total cavity length $L=22\,\mu$m, spatial variations of the cavity field along the growth direction are neglected within the active region. The centre of this layer is located at  $z_0=12$ $\mu$m away from the metal-dielectric interface ($z=0$) and $8\,\mu$m away from the dielectric-air interface ($z=L$). 

To establish the light-matter interaction Hamiltonian, we start from the model introduced in Ref.~\cite{MOrlita:NatPhys2014}, originally formulated for $k_z=0$, and extend it to arbitrary $k_z$ to capture the bulk character of Kane fermions in the MCT layer. The resulting eigenstates are then used to evaluate the photonic pseudopotential, following the procedure outlined in Ref.~\cite{ODmytruk:PRB2021}.
\begin{figure*}[htpb]
     \centering
         \includegraphics[width=1.0\textwidth]{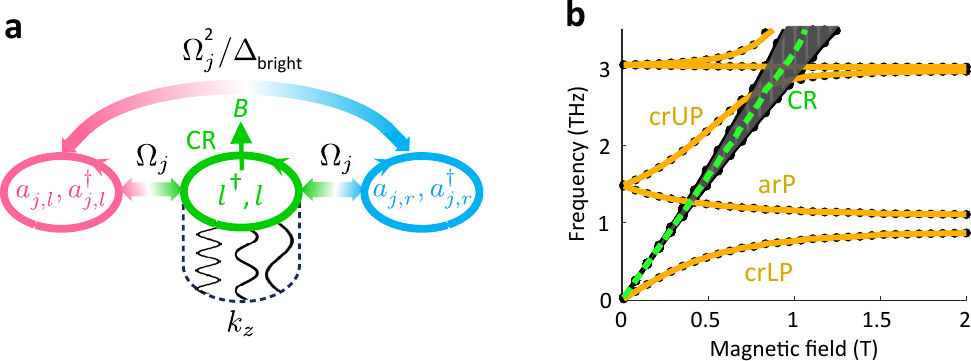}
	\caption{\textbf{Microscopic modelling of MCT polaritons.} (a) Schematic representation of the Hamiltonian~\eqref{H_total_expanded3}. The CR collective ``bright'' mode arises as a coherent superposition of transitions from the highest occupied LL to the lowest unoccupied level, contributed by many transverse wavevectors $k_{z}$. This mode couples resonantly to the co-rotating photon modes M$_j^0$ ($j=1,2$) with strength $\Omega_j$, described by the operators $\hat{a}^{\vphantom{\dagger}}_{j,l}$ and $\hat{a}^\dagger_{j,l}$, and couples antiresonantly to anti-rotating modes associated with $\hat{a}^{\vphantom{\dagger}}_{j,r}$ and $\hat{a}^{\dagger}_{j,r}$. The Hamiltonian further incorporates a diamagnetic $\mathbf{A}^2$ term, which renormalises the bare mode frequencies $\omega_j$ and induces coupling between left- and right-circular polarisations.
    (b) Polaritonic eigenmodes of the full Hamiltonian at 90~K, in which each cyclotron transition with frequency $\Delta_{k_z,\sigma}$ (shaded area) is coupled to the cavity modes (black markers), compared with those of the effective Hamiltonian [Eq.~\eqref{H_total_expanded3}] (orange lines). In the effective description, only the collective ``bright'' mode, with frequency $\Delta_{\mathrm{bright}}$ (green dashed line), interacts with the cavity modes.
    }
	\label{fig:sketch_H}
\end{figure*}
The Hamiltonian describing the coupling between the effective CR in MCT and the two cavity modes $M_j^0$ with frequencies $\omega_{j}$ ($j=1,2$) reads (see Supplementary Information Sec.\ S1):
\begin{align} 
\hat{H} &= \sum_{j} \hbar \omega_{j} \left(\hat{a}^{\dagger}_{j,l} \hat{a}^{\vphantom{\dagger}}_{j,l} + \hat{a}^{\dagger}_{j,r} \hat{a}^{\vphantom{\dagger}}_{j,r} \right) + \sum_{j,j'} \frac{\hbar^2\Omega_{j}\Omega_{j'}}{\Delta_{\textrm{bright}}} \left(\hat{a}^{\vphantom{\dagger}}_{j,r} + \hat{a}^{\dagger}_{j,l} \right)\left(\hat{a}^{\vphantom{\dagger}}_{j',l} + \hat{a}^{\dagger}_{j',r} \right)  \nonumber \\
& + \Delta_{\textrm{bright}} \hat{l}^{\dagger} \hat{l} + \sum_{j} \hbar\Omega_{j} \left[\left(\hat{l}^{\dagger} \hat{a}^{\vphantom{\dagger}}_{j,l} + \hat{l} \hat{a}^{\dagger}_{j,l} \right) + \left(\hat{l} \hat{a}^{\vphantom{\dagger}}_{j,r} + \hat{l}^{\dagger} \hat{a}^{\dagger}_{j,r}  \right)\right],
\label{H_total_expanded3}
\end{align}
where $\hat{a}^{\dagger}_{j,l}$ ($\hat{a}^{\dagger}_{j,r}$) creates a left- (right-) circularly polarized in-plane photon in mode M$_{j}^0$, corresponding respectively to the co-rotating (anti-rotating) mode with respect to the CR for $B>0$ (see Fig.~\ref{fig:sketch_H}a). The CR operator $\hat{l}^{\dagger}$ creates a collective ``bright mode'', a coherent superposition of dipolar excitations from the highest occupied LL to the lowest empty one, $\hat{l}^{\dagger} = \frac{1}{\sqrt{\overline{\nu}N}} \sum_{k_{x},k_{z},\sigma} 
\hat{c}^{\dagger}_{\nu_{\sigma}+1,k_{x},k_{z},\sigma} 
\hat{c}^{\vphantom{\dagger}}_{\nu_{\sigma},k_{x},k_{z},\sigma}$, where $\hat{c}$ and $\hat{c}^{\dagger}$ are the annihilation and creation electron operators. Here we use the Landau gauge where $k_{x}$ is the guiding centre of the cyclotron orbit in the $x$-direction \cite{landau_lifshitz_qm}. The operators $l$ and $l^\dagger$ are quasi-bosonic in the low excitation limit \cite{DHagenmuller:PRB2010}.

The bright-mode energy is defined as $\Delta_{\mathrm{bright}} = \frac{1}{2\mathcal{N}} \sum_{k_z,\sigma} \Delta_{k_z,\sigma}$, corresponding to the average of the cyclotron gaps over all $k_z$ and spin indices $\sigma$, with $\mathcal{N}$ denoting the number of $k_z$ values. As illustrated in Fig.~\ref{fig:sketch_H}b, the polariton frequencies obtained from the diagonalization of the full Hamiltonian (black markers, see Supplementary Information for details), which incorporates all possible cyclotron transitions with energies $\Delta_{k_z,\sigma}$ (shaded regions in Fig.~\ref{fig:fig11}b), are numerically indistinguishable from those derived from the effective model in Eq.~\eqref{H_total_expanded3} (blue curves; also shown as thin black lines in Fig.~\ref{fig:B5T}a,b). Thus, despite the strong frequency spread of the cyclotron gaps shown in Fig.~\ref{fig:fig11}b, the polariton eigenfrequencies of the full Hamiltonian all lie within the values predicted by the effective Hamiltonian \eqref{H_total_expanded3} with a single bright mode concentrating all the oscillator strength. The bright mode co-rotates with the cyclotron motion and, as apparent from Eq.~\eqref{H_total_expanded3}, couples resonantly to left-circularly polarised photons. At the same time, the coupling to right-circular polarisation arises only from anti-rotating terms~\cite{LXinwei:NatPhot2018}, yielding a strong vacuum Bloch-Siegert shift in our data. Besides the bright mode, the full Hamiltonian predicts a large number of dark modes whose energies remain close to those of the bare cyclotron transitions and which are only weakly coupled to the cavity field. We believe that these dark modes could lead to nonlocal effects for resonators featuring strong spatial inhomogeneity of the electric field~\cite{SRajabali:Natphot2021}.

The light-matter coupling frequency reads:
\begin{align}
\Omega_{j} = \Delta_{\textrm{bright}} \sin \left[\frac{(2j-1) \pi z_{0}}{2 L} \right] \sqrt{\frac{4\alpha \,\overline{\nu}}{\pi (2j-1) \sqrt{\epsilon}}},
\label{lightmatterc}
\end{align}
with $\epsilon\simeq13.7$ the background dielectric constant of the MCT layer, $\alpha\simeq1/137$ the fine-structure constant, and $\overline{\nu}=\sum_{k_{z},\sigma}\nu_{\sigma}(k_{z})\,\xi_{\nu_{\sigma},k_{z},\sigma}$ an effective filling factor. The dimensionless factor $\xi_{\nu_{\sigma},k_{z},\sigma}$ accounts for the overlap between LL wavefunctions and is defined in the Supplementary Information. The dependence of the light-matter coupling strength on the square root of the filling factor reflects the cooperative nature of the light-matter coupling~\cite{DHagenmuller:PRB2010,GScalari:Science2012}. The electron density $\rho$ constitutes the only free parameter of the microscopic model. It is determined by fitting the $B=0$ data, namely by matching the M$_1$ resonance computed from the microscopic model to the experimentally observed reflectance minima. The same value for the density is used for transfer matrix simulations (Fig. \ref{fig:B5T}b), which provide excellent agreement with the experimental data. The pertinence of the gyroscopic dielectric tensor used in the transfer matrix method is actually justified by the emergence of the single bright cyclotron resonance as described above.  

Remarkably, the microscopic model at zero magnetic field explains the blue shift of the M$_1$ mode with temperature (Fig.~\ref{fig:fig2}b). In the microscopic model, this shift originates from the diamagnetic $\mathbf{A}^2$ term in Eq.~\eqref{H_total_expanded3}, which saturates to a finite value in the low-field limit, while the light-matter coupling $\Omega_j$ vanishes. This manifests as the screening effect shown in cavity field distribution maps in Fig. \ref{fig:fig2}c.  

Importantly, accounting for the finite thickness of the MCT layer is essential, implying that the out-of-plane momenta $k_{z}$ must be treated as a discrete variable. Taking $k_{z}=2\pi n_{z}/h$, with integer $n_{z}$, we find from Eq.~\eqref{lightmatterc} that the light-matter coupling scales as $\Omega_{j}\!\sim\!\sqrt{h}$ (for large $h$), reflecting its dependence on the electronic density. This indicates that even larger coupling strengths can be achieved for thick layers that have strong geometrical overlap with the cavity electromagnetic field. 

\begin{figure*}[ht!]
     \centering
         \includegraphics[width=1.0\textwidth]{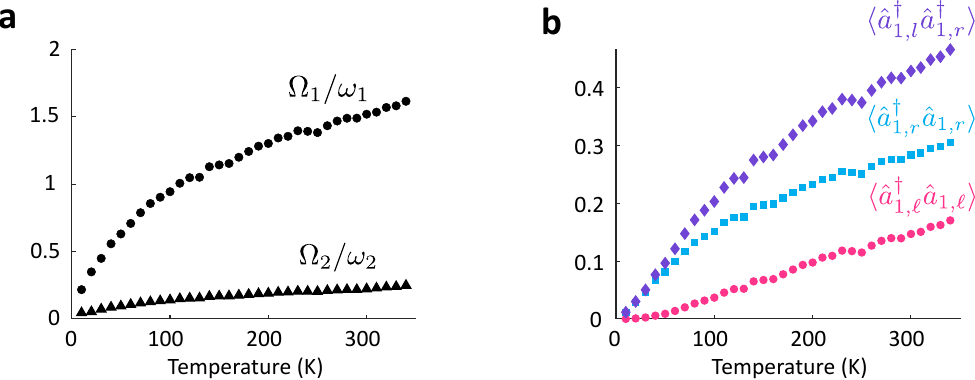}
	\caption{\textbf{Deep strong coupling figures of merit for MCT Landau polaritons}. (a) Temperature dependence of the normalized coupling strength $\Omega_{j}/\omega_{j}$ for the photon modes M$_j^0$ ($j=1,2$). (b) Ground-state photon populations in the first cavity mode M$_1^0$, for left- (pink) and right- (blue) circular polarisations, $\langle a^{\dagger}_{1,p} a^{\vphantom{\dagger}}_{1,p} \rangle$ ($p=l,r$), together with the inter-polarization correlations $\langle a^{\dagger}_{1,l} a^{\dagger}_{1,r} \rangle$. Here $\langle \cdots \rangle$ denotes the expectation value evaluated in the ground state of the Hamiltonian Eq.~\eqref{H_total_expanded3}.
    }
	\label{fig:100K}
\end{figure*}


We are now in a position to estimate the normalised coupling strength $\Omega_{j}/\omega_{j}$---the relevant figure of merit for the ultrastrong and deep-strong-coupling regimes---as a function of temperature (or equivalently carrier density), as shown in Figure~\ref{fig:100K}a. We find that $\Omega_{1}/\omega_{1} > 1.5$ at room temperature, placing the system in an unprecedented deep-strong-coupling regime, while even the second cavity mode at $\omega_{2}/(2\pi)=3$~THz reaches $\Omega_{2}/\omega_{2}\gtrsim 0.2$, well within the ultrastrong coupling regime. Another key figure of merit in these extreme light-matter interaction regimes is the emergence of anomalous correlations in the polaritonic ground state (see Supplementary Information for details). As shown in Fig.~\ref{fig:100K}b, we identify three correlations that acquire the most significant amplitudes: $\langle \hat{a}^{\dagger}_{1,l} \hat{a}^{\vphantom{\dagger}}_{1,l} \rangle$ and $\langle \hat{a}^{\dagger}_{1,r} \hat{a}^{\vphantom{\dagger}}_{1,r} \rangle$, corresponding to the populations of left- and right-circularly polarised photons, respectively, and $\langle \hat{a}^{\dagger}_{1,l} \hat{a}^{\dagger}_{1,r} \rangle$ (together with its complex conjugate), which describes photon pairs with opposite circular polarisations. The latter originates from the diamagnetic term in the Hamiltonian Eq.~\eqref{H_total_expanded3}, which induces the anti-resonant coupling between left- and right-circular polarisations, and provides a clear signature of strong squeezing in the polaritonic ground state~\cite{Ciuti2005}. The order-unity values reported in Fig.~\ref{fig:100K}b are remarkable for a single-mode system and constitute further evidence of the deep-strong-coupling regime reported here~\cite{Mornhinweg2023}.

\section{No-go theorem for Kane fermions}

Finally, our work settles the long-standing controversy regarding the possible existence of the Dicke-like superradiant phase transition for materials described by Dirac-like low-energy models coupled to a cavity field.

More than a decade ago, it was pointed out that the light-matter Hamiltonian for Dirac fermions in graphene does not contain any $\mathbf{A}^2$ term when the minimal substitution $\mathbf{p}\rightarrow(\mathbf{p}-e\mathbf{A})$ is applied directly to the low-energy Dirac Hamiltonian, which is linear in momentum~\cite{DHagenmuller:PRL2012}. This has led the authors of Ref.~\cite{DHagenmuller:PRL2012} to the conclusion that graphene coupled to a cavity could therefore exhibit a superradiant phase transition when brought to the ultrastrong coupling regime. Subsequently, however, it was pointed out that the low-energy Dirac description necessarily involves an ultraviolet cutoff, which breaks gauge invariance, and that a dynamically generated $\mathbf{A}^2$ term prevents the phase transition from occurring~\cite {Chirolli:PRL2012}. 

More generally, it has been emphasised that a direct projection of the minimal-coupling Hamiltonian onto the low-energy sector can lead to a breakdown of gauge invariance~\cite{Chirolli:PRL2012,DeBernardis2018,Li2020}, which is precisely the procedure employed in Ref.~\cite{DHagenmuller:PRL2012}. A similar approach applied to the Kane model, which also exhibits a linear momentum dependence, would also yield an effective Hamiltonian missing the $\mathbf{A}^2$ contribution and thus featuring a superradiant phase transition in the ultrastrong coupling regime. Such a scenario is incompatible with our experimental observations, as we do not observe any qualitative change in the polariton dispersion when crossing the critical coupling $\Omega/\omega= 0.5$ predicted by Dicke-like models. 

Instead, it has been proposed~\cite{DiStefano2019,PhysRevA.102.023718,ODmytruk:PRB2021} that gauge invariance can be preserved by first projecting the fermionic fields onto the low-energy sector and subsequently deriving the light--matter coupling term using a consistently projected unitary transformation acting on the Hamiltonian. In the present work, we follow exactly this procedure to derive the theoretical model. The resulting Hamiltonian shows excellent agreement with the experimental data and does not exhibit any critical point. Within Eq.~\eqref{H_total_expanded3}, this behaviour is ensured by the presence of the diamagnetic $\mathbf{A}^2$ contribution, whose strength becomes comparable to, or even larger than, the bare-mode contributions in the deep-strong-coupling regime. Moreover, the Hamiltonian~\eqref{H_total_expanded3} is strictly identical to that used to describe a GaAs quantum well with quadratic momentum dispersion coupled to a THz photonic crystal cavity~\cite{LXinwei:NatPhot2018}. In the single-mode limit, it also reduces exactly to the Hamiltonian derived in Ref.~\cite{Chirolli:PRL2012} for a graphene sheet coupled to a cavity mode. We therefore conclude that procedures ensuring proper low-energy truncations of the Hilbert space for the light--matter coupling of Kane or Dirac fermions yield results that are fully consistent with the experimental observations and confirm the absence of a superradiant phase transition in non-perturbative regimes of cavity QED.

\section{Conclusions and outlook}

We have realised Landau polaritons using the bulk Hg$_{0.85}$Cd$_{0.15}$Te layer in the Fabry-Perot cavity and accessed the deep-strong-coupling regime with Kane fermions. Normalised coupling ratio reached record $\Omega/\omega = 1.6$ at room temperature through thermal excitation of carriers. The temperature evolution of the spectra further reveals how electronic screening reshapes the cavity field and controls the polaritonic response, while the magnetic field provides an additional tuning parameter. Unlike previous platforms relying on subwavelength metamaterial resonators for local field enhancement \cite{GScalari:Science2012, FAppugliese:Science2022, JMornhinweg:PRL2021}, the present architecture exploits the bulk nature of the active medium, such that the collective coupling strength grows with layer thickness, offering a direct route towards even stronger coupling regimes.

To describe polaritons formed from 3D Kane fermions, we developed a gauge-invariant microscopic theory fully accounting for their relativistic energy dispersion. This framework encompasses previous models based on parabolic bands~\cite{DHagenmuller:PRB2010,LXinwei:NatPhot2018} and extends naturally to two-dimensional relativistic systems such as graphene. The quantitative agreement between experiment and theory rules out the superradiant phase transition in these systems, thereby resolving a long-standing controversy. Our microscopic description in terms of a single bright-mode also agrees well with transfer-matrix simulations describing the optical response in terms of an effective dielectric tensor.

Beyond the present realisation, the Fabry-Perot geometry offers a versatile platform to explore multipartite deep-strong-coupling systems, including hybridisation with disparate quasiparticles, such as magnons~\cite{Bialek_2025}. Carrier populations could also be controlled optically rather than thermally, enabling low-temperature operation and time-resolved studies of polariton dynamics with ultrafast THz spectroscopy~\cite{Gunter2009Subcycle, JMornhinweg:PRL2021}. The same geometry may provide access to quantum vacuum fluctuations through THz electro-optical detection~\cite{RiekC.2015Dsoe, Benea-ChelmusIleana-Cristina2019Efcm}, offering a possible route to probe the large and unusual quantum correlations predicted here. More generally, combining this architecture with transport-compatible heterostructures may open opportunities to study electronic transport in the deep-strong-coupling regime~\cite{Appugliese2022Breakdown, GParaviciniBagliani:Natphys2019, JEnknerNat2025}, as well as for tunable cavity-coupled THz detector and emitter \cite{BenhamouBui2026} architectures.

\bmhead*{Supplementary information.}
Supplementary information is available for this paper here.

\bmhead*{Acknowledgments.}
This work was made possible thanks to high-quality MCT samples grown by molecular beam epitaxy by N.~N.~Mikhailov and S.~A.~Dvoretsky (RAS, Novosibirsk) in 2019. We gratefully acknowledge access to this material. 

\section*{Declarations}
\begin{itemize}
\item Funding
\begin{description}
\item The work was supported by the European Union through the ERC-ADVANCED grant TERAPLASM (No. 101053716). Views and opinions expressed are, however, those of the author(s) only and do not necessarily reflect those of the European Union or the European Research Council Executive Agency. Neither the European Union nor the granting authority can be held responsible for them. We also acknowledge the support of the "Centre for Terahertz Research and Applications (CENTERA2)" project (FENG.02.01-IP.05-T004/23) carried out within the "International Research Agendas" program of the Foundation for Polish Science, co-financed by the European Union under European Funds for a Smart Economy Programme. This work was partially supported by Pasific2 of the Polish Academy of Sciences, sponsored by the European Union's Horizon 2020 research and innovation program under the Marie Sklodowska-Curie grant agreement No.847639, and by the Ministry of Education and Science of Poland. This work was also partially supported by the Sonata Bis-13 2023/50/E/ST3/00584 grant of the National Science Centre of Poland. Y. T. acknowledges financial support from ERC-COG-863487 UNIQUE. 
This research was partially supported by the Foundation for Polish Science project ”MagTop” no.\ FENG.02.01-IP.05-0028/23 co-financed by the European Union from the funds of Priority 2 of the European Funds for a Smart Economy Program 2021-2027 (FENG).
\end{description}
\item Conflict of interest/Competing interests 
\begin{description}
\item The authors declare no competing interests.
\end{description}
\item Ethics approval 
\begin{description}
\item Not applicable.
\end{description}
\item Consent to participate
\begin{description}
\item Not applicable.
\end{description}
\item Consent for publication
\begin{description}
\item Not applicable.
\end{description}
\item Data availability
\begin{description}
\item The experimental data and simulation are available u
\end{description}
\item Code availability 
\begin{description}
\item The codes used in the theory part of this study are available from the corresponding author upon reasonable request.
\end{description}
\item Authors' contributions
\begin{description}
\item D.Y. conceived the idea, prepared the samples, performed experiments, discussed microscopic and transfer-matrix models, discussed data interpretation, and contributed to the writing of the manuscript. D.H. developed the microscopic model, performed the corresponding calculations, discussed data interpretation, and contributed to the writing of the manuscript. N.Ch. developed the transfer-matrix model, and contributed to the writing of the manuscript. Y.I. developed the experimental setup, and performed experiments. A.K performed transport measurements. Y.T. discussed microscopic and transfer-matrix models, discussed data interpretation, and contributed to the writing of the manuscript. W.K. supplied resources, and contributed to the writing of the manuscript. M.B. conceived the idea, developed the experimental setup and software, discussed microscopic and transfer-matrix models, discussed data interpretation, and contributed to the writing of the manuscript.
\end{description}
\end{itemize}

\section*{Methods}

\subsection*{Sample preparation}

A bulk MCT crystal is grown by molecular beam epitaxy on a GaAs substrate. The substrate is removed by wet etching in a H$_2$O$_2$/NH$_4$OH solution, yielding a free-standing flake with lateral dimensions of 5~mm~$\times$~5~mm. The resulting stack consists of three main layers: 10~$\mu$m of epoxy glue, 4~$\mu$m of Hg$_{0.85}$Cd$_{0.15}$Te, and a 6~$\mu$m CdTe buffer layer (Fig.~\ref{fig:fig1}a). The central Hg$_{0.85}$Cd$_{0.15}$Te region is symmetrically enclosed by $\sim$1~$\mu$m-thick graded Hg$_{1-x}$Cd$_x$Te layers, in which the Cd content varies monotonically from 45\% to 15\%. Exemplary scanning electron microscopy images of the studied sample are provided in Ref.~\cite{DYavorskiy:JPD2025} and Supplementary Information.

\subsection*{Experimental setup}

The samples are mounted at the centre of a 2~T superconducting magnet and cooled in darkness to 10~K. We measure reflection time-domain traces in the Faraday geometry as a function of magnetic field $B$ (20~mT steps) and temperature $T$ (10~K steps).
Measurements are performed using a TeraFlash pro time-domain spectrometer (\textit{TOPTICA Photonics}) based on photoconductive antenna emitter–detector pairs. Two antenna sets (centred at 0.5 and 1~THz) provided a spectral bandwidth of 0.14–3.5~THz with a 5~GHz resolution (200~ps scan window) and a signal-to-noise ratio exceeding 120~dB.
The THz beam is directed onto the samples at an incidence angle of 8° using off-axis parabolic mirrors. The optical path outside of the cryostat is continuously purged with dry nitrogen.




\subsection*{Transfer matrix model}

We employ the transfer-matrix formalism \cite{Katsidis2002} to quantitatively describe the magneto-optical response of the full heterostructure, including the metallic holder. We consider an electromagnetic plane THz wave incident normally from vacuum. The stack of layers ends with a semi-infinite metallic copper substrate. Material parameters used in the modelling are provided in the Supplementary Information Sec.\ S2.

\subsection*{Microscopic model}

The light-matter coupling is described using a gauge-invariant microscopic quantum model based on the low-energy Kane Hamiltonian for MCT fermions in a magnetic field, extended to finite $k_z$ to account for the bulk character of the electronic states. The first two open Fabry-Perot cavity modes are quantised and coupled to LL transitions through a projected minimal-coupling Hamiltonian derived via a unitary gauge transformation depending on the LL wavefunctions. In the low-excitation regime, the electronic (dipolar) transitions are mapped onto collective bosonic magneto-exciton operators, yielding an effective bright mode concentrating the oscillator strength of the continuum of cyclotron transitions. Spin-flip excitations are shown to contribute negligibly to the collective coupling and are neglected. The resulting Hopfield-like Hamiltonian is diagonalised to obtain the polariton eigenmodes and ground-state correlations used to model the experimental spectra.

\bibliography{biblio}

\end{document}